\PassOptionsToPackage{table,xcdraw}{xcolor}
\documentclass[sigconf, manuscript, preprint]{acmart}

\usepackage{bm}
\usepackage{color}
\usepackage{tabularray}
\usepackage{longtable}
\usepackage{tabularx}
\usepackage{multirow}
\usepackage{subcaption}

\copyrightyear{2024}
\acmYear{2024}
\setcopyright{rightsretained}
\acmConference[IUI '24]{29th International Conference on Intelligent User Interfaces}{March 18--21, 2024}{Greenville, SC, USA}
\acmBooktitle{29th International Conference on Intelligent User Interfaces (IUI '24), March 18--21, 2024, Greenville, SC, USA}
\acmDOI{10.1145/3640543.3645202}
\acmISBN{979-8-4007-0508-3/24/03}

\begin{document}


\title{Impact of Voice Fidelity on Decision Making: A Potential Dark Pattern?}

\author{Mateusz Dubiel}
\authornote{Both authors contributed equally to this research.}
\orcid{0000-0001-8250-3370}
\affiliation{%
  \institution{University of Luxembourg}
  \country{Luxembourg}}
\email{mateusz.dubiel@uni.lu}

\author{Anastasia Sergeeva}
\authornotemark[1]
\orcid{0000-0003-3701-3123}
\affiliation{%
  \institution{University of Luxembourg}
  \country{Luxembourg}}
\email{anastasia.sergeeva@uni.lu}

\author{Luis A. Leiva}
\orcid{0000-0002-5011-1847}
\affiliation{%
  \institution{University of Luxembourg}
  \country{Luxembourg}}
\email{name.surname@uni.lu}


\begin{abstract} 
Manipulative design in user interfaces (conceptualized as dark patterns) has emerged as a significant impediment to the ethical design of technology and a threat to user agency and freedom of choice. While previous research focused on exploring these patterns in the context of graphical user interfaces, the impact of speech has largely been overlooked. We conducted a listening test ($N = 50$) to elicit participants' preferences regarding different synthetic voices that varied in terms of synthesis method (concatenative vs. neural) and prosodic qualities (speech pace and pitch variance), and then evaluated their impact in an online decision-making study ($N = 101$). Our results indicate a significant effect of voice qualities on the participant's choices, independently from the content of the available options.  
Our results also indicate that the voice's perceived engagement, ease of understanding, and domain fit directly translate to its impact on participants' behavior in decision-making tasks.
\end{abstract}



\begin{teaserfigure}
  \includegraphics[width=\textwidth]{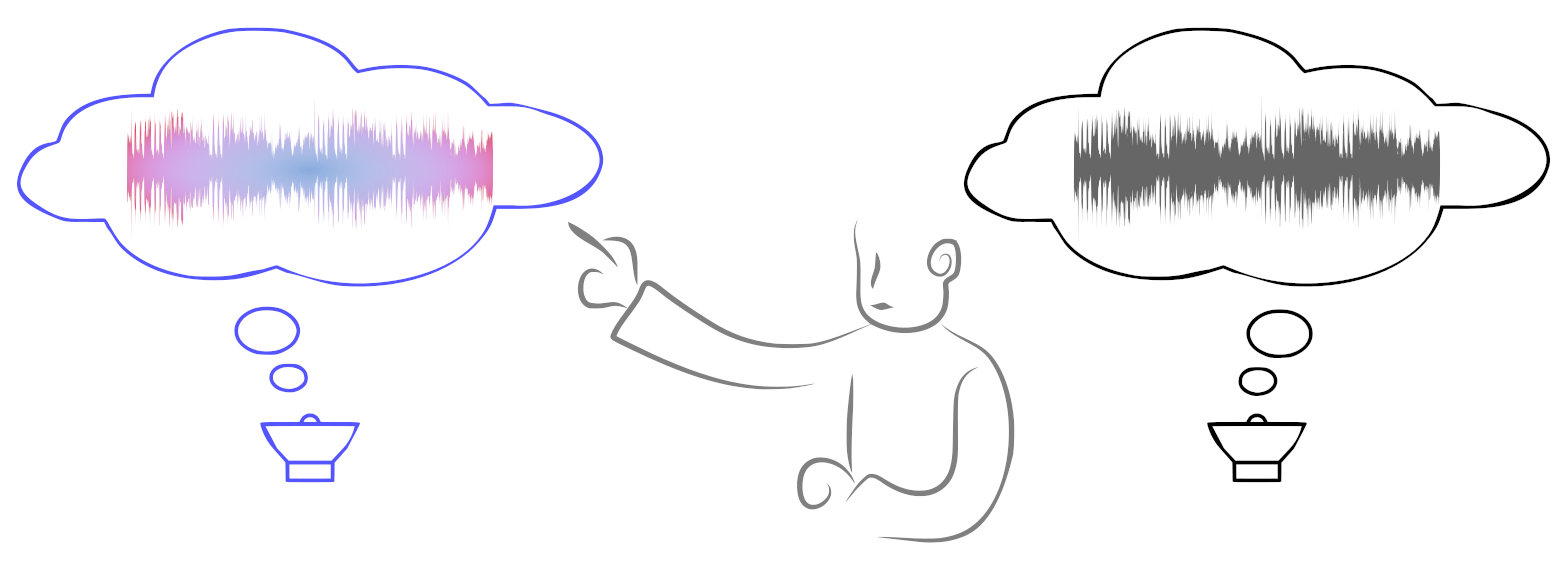}
  \caption{
    Modern synthetic voices may sound very natural to end users; therefore, they have the potential for persuasion and in turn, affect user behaviour.
    We believe that this could be an emerging dark pattern in conversational user interfaces.
  }
  \Description{Abstract representation of interface interference dark patterns in voice technology.}
  \label{fig:teaser}
\end{teaserfigure} 

\maketitle

\section{Introduction}\label{introduction}
Manipulative designs in user interfaces, conceptualised as dark patterns\footnote{We acknowledge the ongoing discussion within the community regarding using the term ``deceptive design patterns'' instead of ``dark patterns''. However, following recent literature~\cite{gray2023towards}, we decided to use the term ``dark patterns'' in the context of this work, since not all designs classified as dark patterns can be explained solely in terms of their deceptive capabilities (cf.~\cite{monge2023defining}).}, have emerged as a significant impediment to the ethical design of technology. They threaten to undermine the values prioritising user preferences, interests, consumer rights, and data protection rights. These patterns manifest themselves as manipulative design decisions, deliberately sidelining the user's best interest to serve the business motives behind them. Numerous academics, practitioners, and supervisory bodies have ventured into classifying these patterns~\cite{gray2018dark,bosch2016tales,jarovsky2022dark}, however, a noticeable limitation in the current discourse is its focus on desktop interface mechanics, such as those seen in various e-commerce~\cite{mathur2019dark} and, to some extent, in broader areas of screen-based mobile~\cite{di2020ui} and desktop interfaces~\cite{bongard2021definitely}. Notably, this current focus largely overlooks the evolving landscape of digital interaction that transcends conventional screen-based mobile and desktop paradigms, such as voice-based interfaces and, more generally, Conversational Agents (CAs).

CAs that operate with a voice interface, such as Amazon Alexa, Apple Siri, or Google Home are becoming increasingly ubiquitous. According to Statista Report, in May 2023 there were over 95 million smart speakers  installed in the United States alone.\footnote{https://www.statista.com/topics/4748/smart-speakers/} While CAs are still predominantly used for simple tasks such as checking the weather, playing music, or setting alarms~\cite{ammari2019music}, a growing number of users is expecting to use them routinely for purchasing products and services online~\cite{nationalpublicmediaSmartAudio}. Consequently, the increase in transaction-oriented interactions provided by the present-day CAs opens up ways for implementing dark patterns that can potentially affect users' decisions towards their disadvantage~\cite{dubiel2022conversational}. While the potential of dark patterns to influence users' interactions with voice-based systems has recently been highlighted by Owens et al.~\cite{owens2022exploring}, this area of research still remains understudied. 

It should be noted that while voice interfaces share the manipulative potential akin to their text-based counterparts, they introduce a unique set of challenges and considerations. A critical element to consider is the prosodic aspect of communication, which, among the other characteristics, pertains to the rhythm, intonation, and stress patterns of speech~\cite{belin2011understanding}. Based on the assumptions of deceptive design experts, this can influence how different options are presented to users by voice assistants in decision making scenarios~\cite{owens2022exploring}. Still, the effect's exact size and the extent to which the effect can be hidden from the user, and therefore considered as manipulative, remains unknown. With this paper, we aim to bridge this knowledge gap. We endeavour to spotlight the largely unexplored territory of prosodic manipulation, by investigating if voice assistants can be used to steer user choices subtly. We also investigate, how users perceive the effect of the voice assistant presentation on their choices. Specifically, we seek to answer the following two research questions. 

\begin{itemize}
    \item \textbf{RQ1:} To what extent the presentation of options provided with synthetic voices can affect users' choices?
    \item \textbf{RQ2:} How impactful do users consider the role of voice in affecting their choices? 
\end{itemize}

Overall, this paper makes the following contributions: 

\begin{enumerate}
    \item We provide evidence for a link between perceived voice characteristics and its impact on user choices in a decision-making scenario.
    \item We showcase that the type of voice may affect participants' decision without them being aware.
    \item We consider the ethical implications of our findings and make recommendations on how designers and policy makers can ensure that CAs are designed to benefit users and promote their agency. 
\end{enumerate}

\section{Background and Related Work}\label{background}
Choice architecture~\cite{thaler2009nudge} relies on designing choice situations where decision-makers are `nudged' towards more beneficial options~\cite{szaszi2018systematic}. As postulated by Thaler and Sunstein, Nudge theory offers less invasive and more subtle ways to influence human behaviour compared to direct interventions~\cite{thaler2009nudge}. It has been applied in several domains such as finance~\cite{benartzi2017should}, health~\cite{dai2021behavioural}, and sustainability~\cite{mont2014nudging} by changing choice defaults and or proving users with social comparisons~\cite{landais2020choice}. It has been also argued that industry practitioners and policy makers can use Nudge theory to arrange decision making context in order to influence users' choices in a cheap and effective way~\cite{hansen2013nudge}. However, despite the potential to inspire positive behavioural change, choice architecture also has a darker side. In this section we will discus both positive and negative implications of nudging and comment on the potential role of speech in influencing human behaviour. 

\subsection{Nudges and Dark Patterns in User Interfaces}
Interdisciplinary research from computer science, economy, law, and psychology, among other fields, frequently examines how certain subtle techniques can influence the presentation of choices to the end-user. 
While some argue that directing user choices can be beneficial when done in the user's interest~\cite{thaler2009nudge}, increasing attention is being given to instances where these techniques serve the business stakeholders' interests instead~\cite{hornuf2022digital}. Recent studies also highlighted a number of instances where these techniques can steer people's decisions in specific directions (e.g., choosing a subscription option, which is more profitable for the company, or sharing an extensive amount of personal data~\cite{mathur2021makes}. 

In the field of Human-Computer Interaction, the intentional design of interfaces to mislead or manipulate users for business gain or extensive data collection is termed ``dark patterns.'' These are featured in  interfaces purposely designed to confuse users, hinder them from expressing their true preferences, or coerce them into specific actions~\cite{gray2018dark}. Since Brignull introduced the term in 2011, various classifications have been developed to categorise and explain different types of dark patterns and their interrelationships~\cite{gray2023towards, gray2023mapping}. 

Some classifications focus on the specific harms that dark patterns can cause, such as privacy issues~\cite{bosch2016tales, jarovsky2022dark}. Others highlight the broad principles that make these patterns effective~\cite{mathur2021makes, gray2018dark}. Several studies investigate dark patterns in specific interactions such as gaming~\cite{zagal2013dark}, online shopping~\cite{mathur2019dark,tuncer2023running}, and video streaming~\cite{chaudhary2022you}. There is also research dedicated to analysing the differences in dark patterns on different modalities, such as mobile versus desktop devices~\cite{di2020ui,gunawan2021comparative}. Experimental studies on dark patterns have revealed that users more easily recognise some practices (e.g., fake urgency presented via countdown timers) than others~\cite{bongard2021definitely}. Moreover, subtler manipulations often yield better results for companies~\cite{luguri2021shining}. Several legislative and customer protection bodies (e.g., \textit{European Commission, Bureau Européen des Unions de Consommateurs}, etc.)  have recently suggested measures to address the most overt and deceptive dark patterns~\cite{CNIL2019, BEUC2022}. While such legislation may offer a better protection for users from the most prevalent dark patterns, it should be noted that there are new strands of dark patterns that are constantly emerging that may be more difficult to identify for both users and regulators. 

\subsubsection {Dark Patterns Beyond Screen-based Interfaces}
Until now,  only a few studies have focused on potential dark patterns beyond  e-commerce and gaming applications. For example, Kowalczyk et al.~\cite{kowalczyk2023understanding} studied dark patterns in Internet of Things (IoT) devices, finding not only a high number of already identified dark patterns (e.g., hidden subscription, bait and switch, obfuscation, etc.) but also unveiling new types specific to IoT devices, such as `pay for the long term use'. Wang et al.~\cite{wang2023dark} delved into manipulative designs in Augmented Reality and found that manipulations involving lighting and object interference impact participants' responses. They also observed that haptic feedback can guide users in a specific direction therefore unconsciously manipulating their choice. Focusing on design aesthetics, Lacey and Caudwell~\cite{lacey2019cuteness} explored the concept of ``cuteness'' (considered as visual appeal) as a dark pattern in the home robot design, which creates an affective response in the user for the purpose of collecting emotional data.
Lastly, Owens et al.~\cite{owens2022exploring} sought expert opinions on potential dark patterns in voice-based interfaces, focusing on a range of problematic scenarios, which include both interaction parameters of voice assistant technology and speech properties such as volume, pitch, rate, fluency, pronunciation, articulation to emphasise certain options and, consequently, increase their prominence to the user. Similarly, Dula et al.~\cite{dula2023identifying} discussed the parameters of the voice as a part of ``dishonest anthropomorphism'', which can be viewed as a deceptive design feature, whereby the human-likeness of the agent is being used to influence the users. 

\subsubsection {Benevolent Nudges in Voice Interfaces}
At the same time, several studies discussed the positive aspect of nudging users via voice-based interfaces. For example, Gohsen et al.~\cite{gohsen2023guiding} discussed how to use nudges to support information-seeking behaviour via voice-assistant-based information search.
Tussyadiah and Miller~\cite{tussyadiah2019nudged} discussed the possibility of leveraging pro-environmental behaviour via nudging by voice-assistant. Similarly, He and Jazizadeh~\cite{he2022nudging} discussed nudging for Energy-Saving behaviour. Studies also showed that in certain scenarios some people are open towards the possibility of voice assistant nudging them to improve their negative habits~\cite{volkel2021eliciting}.
However, neither of the presented studies explicitly discussed the characteristics  of voice in relation to its nudging ability. In the following, we will briefly discuss the role of prosody in speaker perception and forming attitudes. 

\subsection{Role of Speech Prosody and Pace in Voice Perception and User Behaviour}
Voice is considered as one the main sources of information that shape our social impression of other people~\cite{belin2011understanding}. Specifically, non-verbal cues such as intonation, emphasis and rhythm influence both our perception of a speaker~\cite{varghese2020you} and cognitive processing of information~\cite{rodero2016influence}.  While human voice is characterised by a wide range of prosodic aspects, the mean fundamental frequency (F0), judged as voice pitch~\cite{titze1998principles}, is considered to play a central role in social judgements of human voices, especially among males~\cite{schild2020linking}. 
Research on voice perception has linked lower pitch to higher levels of attractiveness~\cite{puts2016sexual}, competence~\cite{oleszkiewicz2017voice}, and dominance~\cite{vukovic2011variation}. When it comes to speech pace, moderate speech of around 180 words per minute has been found to be optimal for recall and recognition of information~\cite{rodero2016influence}. In general, speech needs to be fast enough to attract listeners' attention and moderate so that it does not hinder comprehension~\cite{rodero2022expressive}. However, while the relationship between fast-to-moderate speech, lower mean F0 and higher attractiveness and dominance in men has been found quite consistently across several studies~\cite{puts2016sexual, oleszkiewicz2017voice, vukovic2011variation}, evidence of the relationship between the voice perception and its impact on user behaviour in decision making tasks is unclear~\cite{dubiel2020persuasive}. 

\subsubsection{Affect Heuristics and User Behaviour}
According to Slovic~\cite{slovic2007affect}, people determine their attitudes by consulting their feelings, 
and involve emotions when making their judgements. Specifically, he postulated a link between affect and cognition and forming options and attitudes, where `people consult or refer to ``affect pool'' containing all the positive and negative tags consciously or unconsciously associated with the representations'~\cite{slovic2007affect}. By extension, Kahneman et al.~\cite{kahneman2021noise} posit that affective impressions that are readily available and require less mental resources can bias human decision making and affect behavioural outcomes. In this paper, we set out to explore the link between perceived affect of voice and its impact on users in a decision making scenario in a context of `food inspirations'. Specifically, we explore how appealing people find different food items when they are presented by different synthetic voices. Drawing from literature in psychology and economics, here, we consider `food inspirations' as a low-involvement scenario. 

\subsubsection{Low-involvement Decision Making}
Research indicates that people rarely engage in an extensive decision‐making process or in-depth evaluation of product features when buying food~\cite{tarkiainen2009product,yeo2017consumer}. From the economics perspective, purchasing food can be generally considered as a low-involvement decision making scenario, since it tends to be more habitual and requires less deliberation compared to high-involvement purchases such as renting a car or buying an insurance policy~\cite{moriuchi2019okay}. Tassielo et al.~\cite{tassiello2021alexa} explored use of CAs for low- and high-involvement decision making and found that users felt more empowered when presented with low-involvement product choices which consequently led to more willingness to make a purchase. Following this line of reasoning, in our study we assume that users will be more likely to be influenced by a CA in scenarios where the decisions they have to make are considered to be `low-involvement'. 

\subsubsection{Synthetic Speech in Conversational Agents}
\label{low_decisions} 
A Text-to-Speech (TTS) system, generally referred to as synthetic speech, converts text into speech \cite{taylor2009text} is an integral part of CAs. While CAs are becoming more frequently used in transactional scenarios,  research indicates that monotonous, robotic and unnatural vocal features of TTS negatively impact users' engagement ~\cite{cambre2020choice,dubiel2020persuasive,choi2020nobody}. This can be addressed by appropriate interaction design  and providing a better fit of voice to application domains by diversifying voices to improve user experience~\cite{sutton2019voice,aylett2014none,motalebi2019can,zargham2021multi,aylett2020voice,laban2022newspod}. Indeed, empirical evidence suggests that voices that are enjoyable to listen to can not only delight users but also establish lasting relationship and long-term usage~\cite{walter2011designing}. Recent research emphasises the importance of voice characteristics such as pace of speech, tone and accent in engaging uses in enjoyable CA experiences~\cite{dubiel2020persuasive,choi2020nobody,shin2020designing}. 

In a recent study Do et al.~\cite{do2022new} have investigated the impact of type of synthetic voice (concatenative vs. neural) on social perception of virtual agents and their persuasiveness. They found that a standard synthetic voice (build using standard concatenative approach) was perceived as more trustworthy than a deep-learning-developed, neural voice which mimics speakers characteristics with high fidelity. In the current study, we compare standard and neural synthetic voices in terms of their suitability for a `food inspiration' agent (i.e., an agent providing participants examples meals that they can cook or order) and explore their potential to affect user choices. Contrary to Do et al.~\cite{do2022new} we focus exclusively on the voice domain and investigate only male voices, to reduce the complexity of the study since the gender of the speaker is known to influence listeners' perceptions of the agent~\cite{mullennix2003social}. Instead, we leave exploration of other voice genders to future work. 

\section{Study}\label{methodology}
Our main research goal is to evaluate the role of synthetic voice in a decision-making task, 
to explore if there is a relationship between perceived qualities of voice and its impact on user's behaviour.
For this, we designed a study comprising two stages, illustrated in \autoref{fig:overview}: 
(1)~Voice Perception stage, where three voices of different prosodic qualities (presented in \autoref{fig:pros}) are evaluated via a listening test, 
and (2)~User Behaviour stage, where the impact of voice is evaluated in a decision-making task. 
Our study was approved by the Ethics Review Panel
of the University of Luxembourg with the ID: ERP 22-005 DPVADM. 

\subsection{Materials}

\subsubsection{Synthetic Voices}
We selected Amazon Polly's American male voice \textit{Joey}\footnote{https://docs.aws.amazon.com/polly/latest/dg/what-is.html} (Standard TTS) as our baseline because it is a popular, high-quality concatenative synthetic voice that is frequently used in voice-over applications. We have also selected two American male neural TTS voices from the TorToiSe repository,\footnote{https://github.com/neonbjb/tortoise-tts} \textit{Les} (Neural TTS1) and \textit{William} (Neural TTS2) as our upper-bound candidates. We decided to use TorToiSe TTS~\cite{betker2023better} since the software is capable of creating highly expressive and natural sounding voices and is publicly available as open source. As a caveat, it should be noted that while TorToiSe neural voices can capture human vocal qualities with extremely high fidelity, they suffer from slow synthesis time, which currently prohibits their commercial implementation. However, we assumed that due to their antropomorthic features, the neural voices will be considered more pleasant to listen to and, in turn, incite a more positive user sentiment~\cite{maharjan2022s}.

To explore the potential of voice to impact participants' behaviour, as explained in \autoref{low_decisions}, we decided to evaluate our voices in a `food inspirations' context. We consider this context as low-involvement, since many food selection decisions are made without much cognitive effort~\cite{daniel2017thinking}). A similar scenario was previously used by Dubiel et al.~\cite{dubiel2023you} in a study which focused on CA's feedback appropriateness.

\begin{figure*}[h!]
    \centering
    \includegraphics[width=0.95\linewidth]{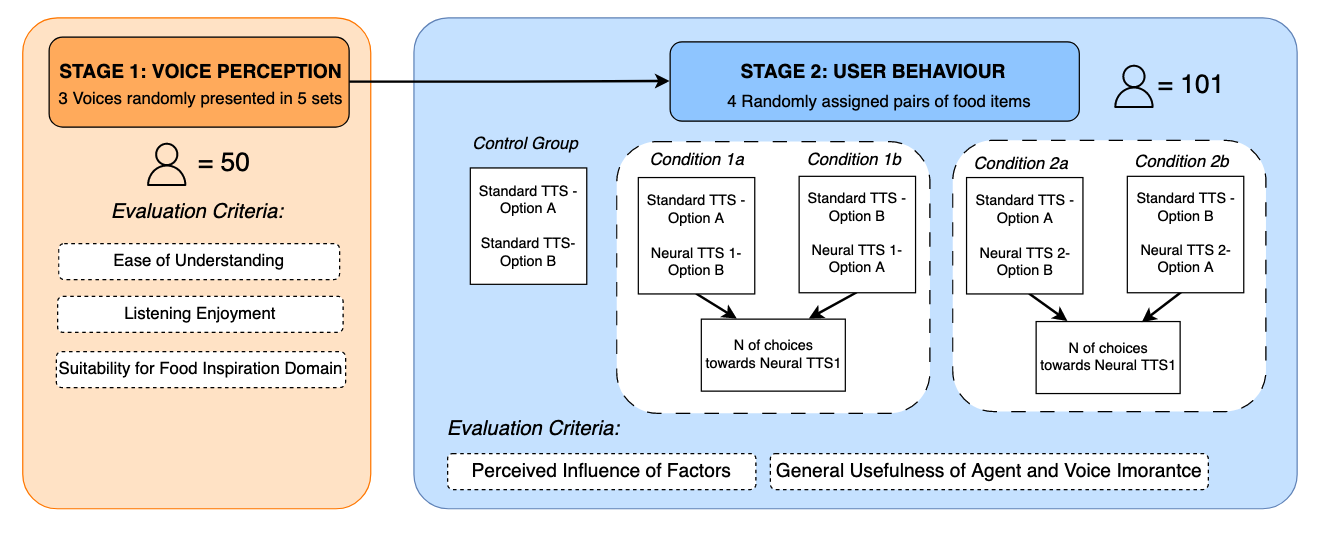}
    \caption{Overview of Experimental Stages.}
    \label{fig:overview}
\end{figure*}

\begin{figure}[h!]
    \centering
    \includegraphics[width=0.65\linewidth]{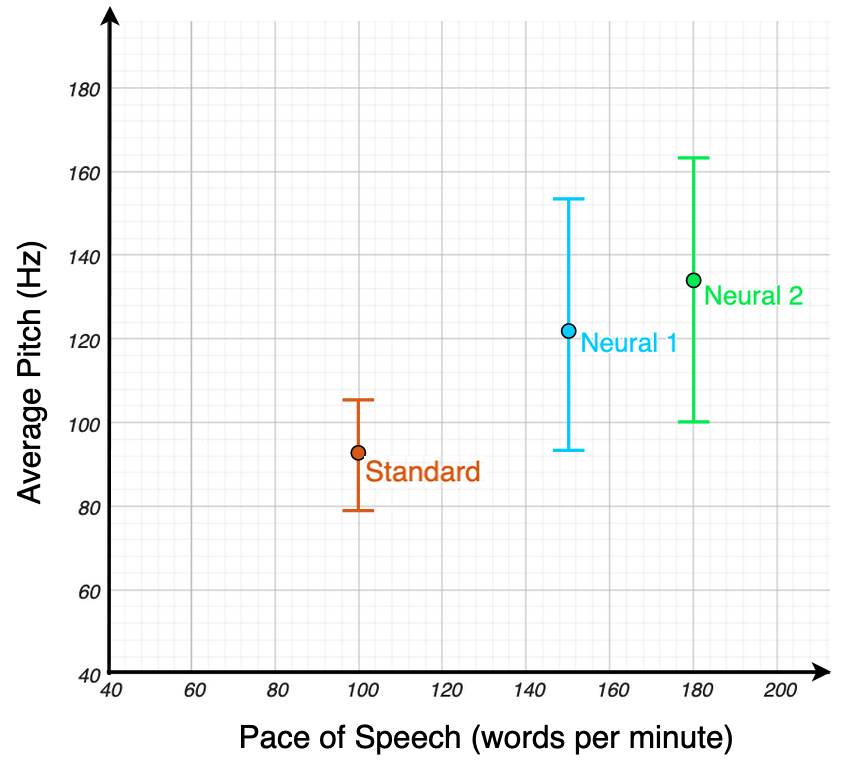}
    \caption{Prosodic qualities of selected voices: Standard TTS (baseline) and two Neural TTS voices. Error bars denote the standard deviation of pitch, based on acoustic analysis.}
    \label{fig:pros}
\end{figure}

As illustrated in \autoref{fig:pros}, we can see that Standard TTS has a lower average pitch and narrower pitch variance (93\,Hz, SD=13\,Hz) compared to Neural TTS1 (122\,Hz, SD=34\,Hz) and Neural TTS2 (133\,Hz, SD=33\,Hz). It is also slower (100 words per minute) compared to Neural TTS1 and Neural TTS2 (150\,WPM and 180\,WPM, respectively). 

\subsubsection{Prompt Generation}
The prompts used in both stages of this study (i.e., Voice Perception and User Behaviour) 
were generated with OpenAI's ChatGPT~\cite{stiennon2020learning}. We used the following strategy to generate our pool of prompts.
First, we determined the high-level categories (e.g., breakfast, pasta, snacks) in which meals are likely to be described based on~\cite{korpusik2019deep}. Second, we specified the parameters of the generated outputs. Our goal was to generate paired meal options that we could use in the User Behaviour stage of the experiment. We also intended to use the pool of generated sentences as material for the Voice Perception stage of the experiment to assess participants' listening experience and determine suitability of the voices for the `food inspirations' domain.

Concerned that the content of options might introduce a bias towards specific choices, we aimed to create pairs that were equivalent in terms of major dietary restrictions (e.g., in one pair of choices, both options would either be vegetarian or non-vegetarian) and perceived health output (we either used or did not use the term ``healthy options'' to indicate a trend toward nutrition and a low-calorie intake). We also experimented with the length of the prompts, ensuring they did not overload users' working memory~\cite{baddeley1992working}. 
Most of the generated prompts we picked for a study were within a 10-word range.
The final format of our prompts is as follows:

\texttt {Generate two options of [healthy/][vegetarian/non-vegetarian][type of the meal], no longer than 12 words [specific requirements, connected with the proposed type of meal, e.g., ``both including vegetables'' for lunch options or ``both including fruits'' for breakfast options].}

The pool of 10 pairs of prompts was synthesised into audio samples. We conducted an internal listening test to eliminate voice samples that included mispronunciations, and artifacts such as clicks and phase inconsistencies. Based on the pre-test, we selected four pairs of options for the behavioural experiment. We have also chosen five single options that were featured in the Voice Perception stage.
The full set of final prompts is provided in \autoref{appendix}. 
The generated voice examples are provided in the Supplementary Materials. 

\subsection{Stage 1: Voice Perception}
We ran an online listening test on the crowd-sourcing platform Prolific\footnote{https://www.prolific.com/} by providing $N=50$ participants with a link to LimeSurvey.\footnote{https://www.limesurvey.org/} 
Participants were presented with five groups of test samples in total. For each group, participants had to rate three samples of the same sentence, either generated by three different speakers (see \autoref{fig:pros}). The audio samples were based on Chat-GPT generated prompts (see \autoref{appendix} and Supplementary Materials for details).

Participants were asked to rate each sample according to three criteria, namely: Ease of understanding~(1), Listening enjoyment~(2), and Suitability for `Food inspirations' domain~(3). All of the three criteria were scored on a five-point Likert scale, where 1 was `Strongly Disagree' and 5 was `Strongly Agree'. To avoid ordering effect biases, we randomised the sequence of samples and the order of speakers (i.e., Standard TTS (Baseline), Neural TTS1 and Neural TTS2). Based on pilot experiments, we informed participants that the average task completion time is five minutes. We asked the participants to use headphones and to conduct the study on a desktop computer or a laptop computer. 

In order to ensure that listening tests were performed diligently, we implemented two attention checks that were interleaved with the questions. Attention check were conducted by randomly introducing additional white noise samples.
Moreover, to ensure high-quality responses, during the recruitment we exercised additional precautionary measures. Participants (crowd workers) were only permitted to take the listening test if they: (1)~were based in the United States, (2)~spoke English as their first language, and (3)~had an approval rate above 99\% in Prolific. The above constraints were introduced to reduce the risk of recruiting individuals who would not complete the study up to the required standard.

\subsection{Stage 2: User Behaviour}
We ran another crowdsourcing study with $N=101$ participants through the Prolific platform linked to LimeSurvey. 
Participants from Stage 1 were not allowed to participate to reduce bias that could have been created due to previous exposure to voices used in the experiment. 

This time the voices were evaluated in a decision-making task where participants were asked to select between two meal options.
Participants chose between two meal options across four pairs, guided by the question, ``Which option do you find more appealing?''.
They were randomly allocated into one of three categories (as illustrated in \autoref{fig:overview}). The meal options and voice types were allocated as follows.

\textbf{Control Condition}: Both meal options were voiced by the same Standard TTS.

\textbf{Experimental Conditions}: 
    One option pair was voiced by Standard TTS and Neural TTS1.
    Another option pair was voiced by Standard TTS and Neural TTS2.

Each experimental condition had two \textbf{sub-variations}:
\begin{itemize}
    \item In the first condition, option A was voiced by Standard TTS and option B by one of the two Neural TTS alternatives.
    \item In the second condition, option B was voiced by Standard TTS and option A by one of the two Neural TTS alternatives.
\end{itemize}

To eliminate ordering bias, options A and B in each pair were presented in random order.

\textbf{Post-experimental Questionnaires}:
After the experiment, participants were asked to complete a follow-up questionnaire in which they were asked to rate seven factors that could have potentially affected their choices. The factors were: ingredients, health and nutritional considerations, familiarity with proposed cuisine, previous positive experience with similar dishes, novelty, interesting combination of ingredients, and voice presentation. Each item was ranked on a five-point Likert scale where 1 denotes `Not Influential at all' and 5 stands for `Really Influential'. These factors were generated by Authors 1 and 2 based on features of the food domain ~\cite{furst1996food} and included general characteristics of the meal, preferences, and health considerations. We also included the presentation factor (the way the assistant presents the meal) to analyse, how much participants are taking into account the possible differences in the presentations provided by different voices. The last section of the questionnaire consisted of two questions: the general usefulness of the voice agents (1: Not Useful at all, 5: Really Useful) and the overall importance of the agent's voice for the user's decision-making  (1: Not Important at All, 5: Really Important). The full questionnaire is presented in  \autoref{appendix}.

\section{Results}\label{results}
\subsection{Stage 1: Voice Perception}
\begin{figure*}[h!]
    \centering
    \includegraphics[width=0.95\linewidth]{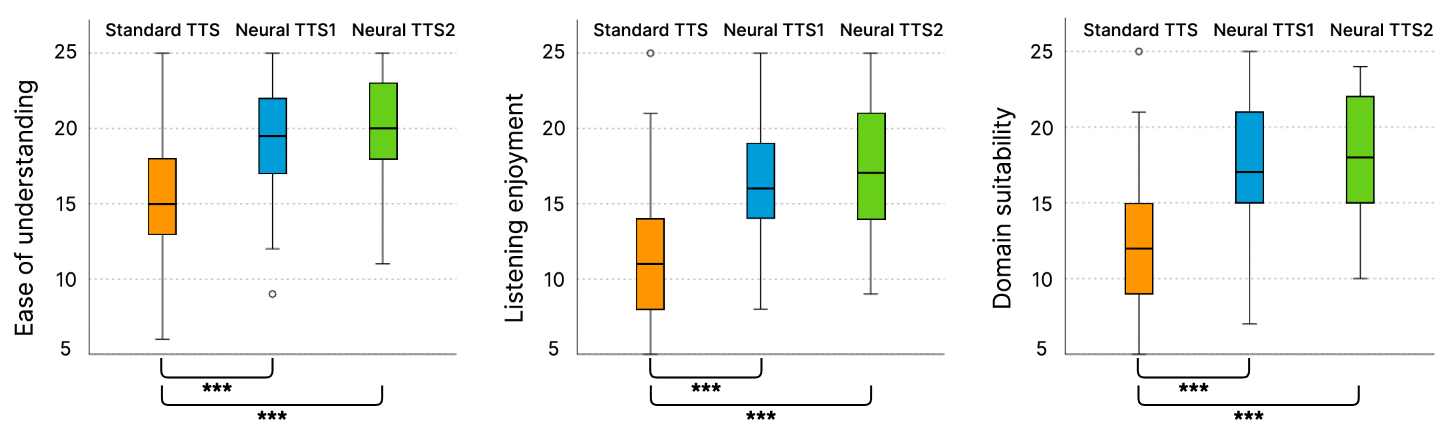}
    \caption{Independent-Samples Kruskal-Wallis ranks for Standard TTS, Neural TTS1, and Neural TTS2
    in terms of: Ease of understanding, Listening enjoyment, and Domain suitability. Note: `***' indicates p < .001}
    \label{fig:pre}
\end{figure*}

Fifty participants took part in the experiment (24 females, 25 males and one undisclosed).
The average age of participants was 36 years (SD = 12.5). The average completion time was 6 mins and 2 secs (SD = 3 mins and 13 secs). The payment for participating in the experiment was \pounds9 per hour. \autoref{fig:pre} presents results of the voice perception part of the study. As can be seen, both neural voices were found to be significantly easier to understand, more enjoyable to listen to, and more suitable for the food inspirations domain.
The correlation analysis using Spearman's \(\rho\) revealed high intercorrelations among the three scales used in the study: 
(1)~\textit{Domain Suitability - Listening Enjoyment} ($\rho(1) = 0.912, p < .001$, 95\% \text{CI} $[0.880, 0.936]$), 
(2)~\textit{Listening Enjoyment - Ease of Understanding} ($\rho(1) = 0.705, p < .001$, 95\% \text{CI} $[0.611, 0.779]$), 
and (3)~\textit{Domain Suitability - Ease of Understanding} ($\rho(1) = 0.750, p < .001$, 95\% \text{CI} $[0.668, 0.814]$).

A non-parametric Kruskal–Wallis one-way analysis of variance between three tested voices (Standard TTS, Neural TTS1, and Neural TTS2) in terms of the three dependent variables (ease of understanding, listening enjoyment, and domain suitability) revealed that Standard TTS scored significantly lower than any of the Neural TTS alternatives.
At the same time, we did not find differences between both Neural TTS voices. 
A summary of these results is presented in \autoref{tab:Krus}.

\begin{table*}[!ht]
\caption{Independent-Samples Kruskal-Wallis Test Summaries. $p$-values are Bonferroni-corrected, to guard against multiple comparisons.}
\begin{tabular}{l *6r}

\toprule
  & \multicolumn{2}{c}{\textbf{Ease of understanding}}
  & \multicolumn{2}{c}{\textbf{Listening Enjoyment}}
  & \multicolumn{2}{c}{\textbf{Domain Suitability}} 
  \\
Omnibus test $\chi^2(2,N=150)$
  & \multicolumn{2}{c}{$\chi^2 = 28.611, p < .001$}
  & \multicolumn{2}{c}{$\chi^2 = 43.219, p < .001$}
  & \multicolumn{2}{c}{$\chi^2 = 43.219, p < .001$} 
  \\
\midrule
  \textbf{Pairwise Comparisons} 
  & \textbf{Std. Test} & \textbf{Adj. $\bm{p}$-value}
  & \textbf{Std. Test} & \textbf{Adj. $\bm{p}$-value}    
  & \textbf{Std. Test} & \textbf{Adj. $\bm{p}$-value}  
  \\
\midrule
Standard TTS - Neural TTS1   & 5.120 & <.001   & 5.488 & <.001   & 5.056 & <.001 \\
Standard TTS - Neural TTS2   & 3.902 & <.001   & 5.879 & <.001   & 5.689 & <.001 \\
Neural TTS1 - Neural TTS2   & 1.218 & .670    & 0.391 & 1.000   & 0.633 & 1.000 \\
\bottomrule
\end{tabular}
\label{tab:Krus}
\end{table*}

We also did not find any statistically significant effects of participants' gender on the perception of any of the voices by any of the tested parameters. 
These results are presented in \autoref{tab:gender}.

\subsection{Stage 2: Voice Impact}

One hundred and one participants took part in the experiment (52 females, 47 males, and 2 undisclosed).
The average age of participants was 44 years (SD = 15.7).  The average completion time was 5\,min and 29\,s (SD = 3\,min and 7\,s). 
The payment for participating in the experiment was \pounds9 per hour. 

\begin{figure}[h!]
    \centering
    \includegraphics[width=0.8\linewidth]{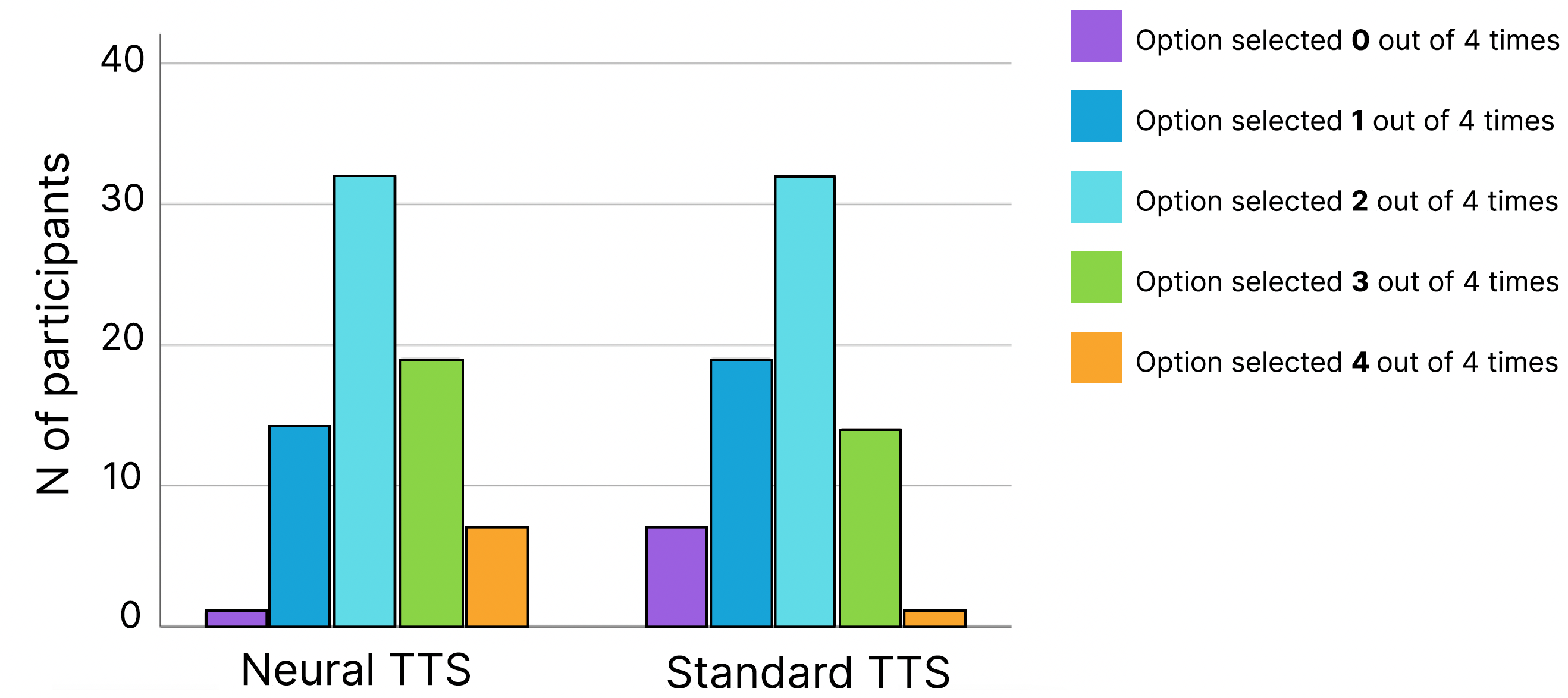}
    \caption{Proportion of selected options when presented by Neural TTS and Standard TTS. 
    For example, the first purple bar on the left means that there were 2 participants who selected options presented by Neural TTS 0 out of 4 times.}
    \label{fig:proaga}
\end{figure}

\subsubsection{User Perception of Voice Assistants' Usefulness in Presenting Options}

We calculated the mean and median from responses to the question regarding the usefulness of voice assistants in providing meal choices. The results yielded a mean value of 3.39 (95\% CI $[3.14, 3.63]$), which is higher than the ``Neutral'' option in the scale. The median was found to be 4. These findings indicate a generally favorable user opinion towards this type of interaction and further support the idea that voice assistant interactions are appropriate for this domain.

\subsubsection{User Behaviour}
\label{sec:behaviour}
To determine if there is a difference in users' choices regarding options provided by the high-fidelity voices (neural) versus the baseline (standard), we first carried out a non-parametric $\chi^2$ test between the sum of choices in favour of option A and in favour of option B regardless of which condition each option was presented. Results indicated that there were no statistically significant differences between the sums of user choices of options A and B in both experimental conditions ($\chi^2(4, 146) = 5.583, p = .233$) and in all samples combined ($\chi^2(4, 202 ) = 6.833, p = .145$).
After that, we conducted a non-parametric $\chi^2$ test between results of choices in favour of and against options, presented by ``high fidelity voices''. The results revealed a significant difference ($\chi^2(4, 146) = 10.515, p = .033$), showing a statistically significant preference for the options presented by high-fidelity voices (see also \autoref{fig:proaga}).
A Cramer’s V analysis revealed a small-to-medium effect size of the differences ($V(1,146) = 0.27$).

A follow-up analysis using the Mann–Whitney U test in the distribution of choices towards options provided by high fidelity voice 
did not yield significant differences between high fidelity voices Neural TTS1 and Neural TTS2 ($U(74) = 768.500, z = .966, p = .334$), 
which is in line with the results observed in Stage 1. 
As in the first study, based on the previous literature suggested that voice perception factors can be connected with gender~\cite{mullennix2003social}, we run an additional $\chi^2$ test to determine if there is a difference in behavioural outcomes for male and female participants. The test did not show statistically significant differences between participants who identified themselves as males or females 
by the number of choices towards Neural TTS ($\chi^2(4, 71) = 2.984, p = .561$).

\subsubsection{User Perception of Voice Affecting their Choices}
\label{sec:choiceperception}
\begin{figure*}[h!]
    \centering
    \includegraphics[width=0.95\linewidth]{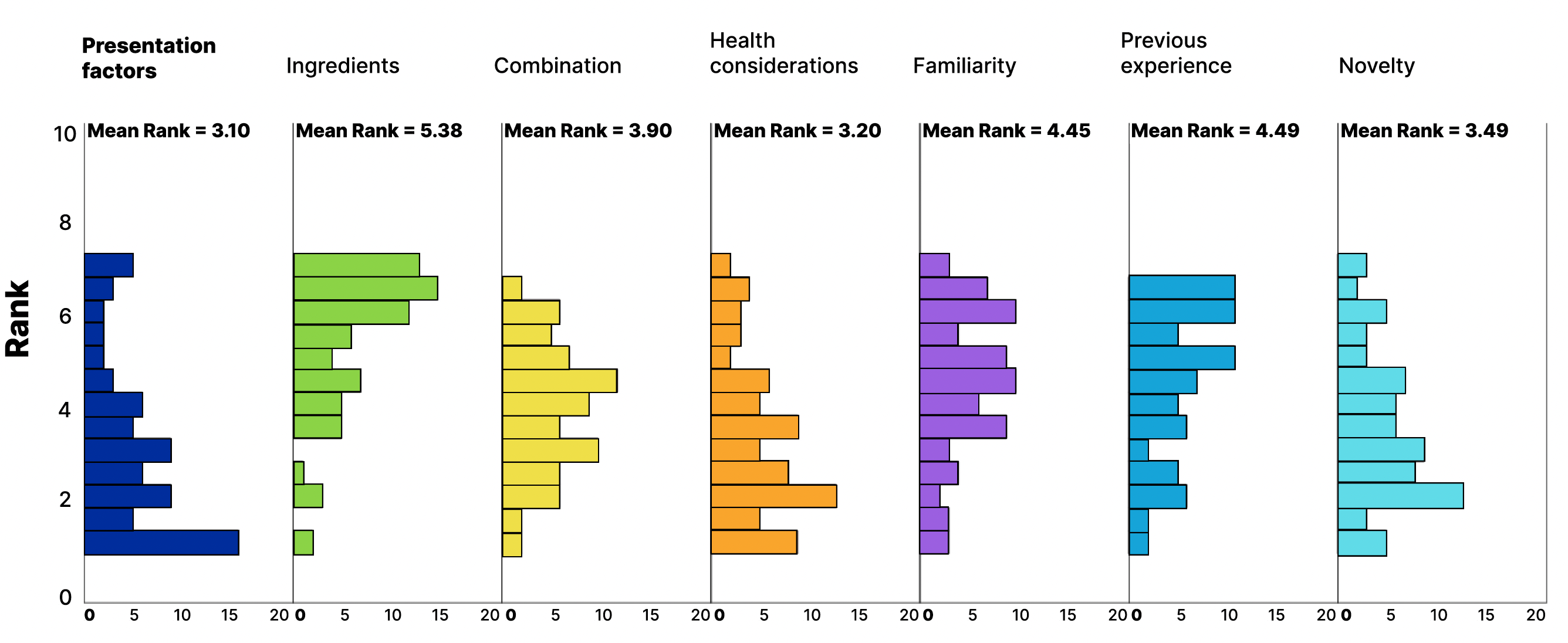}
    \caption{Related-Samples Friedman's Two-Way Analysis of Variance by Ranks in experimental conditions.}
    \label{fig:influ}
\end{figure*}

To understand the role of voice assistant presentation in influencing user choices, we analysed the impact of seven potential factors using the Related-Samples Friedman's Two-Way Analysis of Variance by Ranks in experimental conditions combined. We used the assumption of relatedness of the samples for measuring questions, as all of the samples were presented to participants within the same experimental session. Therefore, we assumed that participants could easily rank the differences between factors importance. The perceived influence of each factor on participants' choices is presented in \autoref{fig:influ}.

\autoref{tab:varbyrank} presents a summary of participants' perceived importance of different aspects of meal presentation. 
We ran pair-wise comparisons (Bonferroni-corrected) and found that presentation factors ranked significantly lower in importance than three out of the six other factors and did not significantly differ from the other three. In fact, it had the lowest mean on rank at 3.10 among all evaluated factors and a raw mean of 2.95 in combined Neural TTS conditions. (This is lower than the ``Neutral'' option in the questionnaire.) 
This suggests that users might not consider the way voice assistants presented the options as a significant factor influencing their decisions. 
The Kruskal-Wallis test also did not reveal significant differences between NeuralTTS1, NeuralTTS2, and StandardTTS 
in the factor ``Presentation Factors'' ($H(2) = 5.883, p = .053$).

If we consider a combined sample of both Neural TTS voices against the Baseline condition, 
the results reveal that in the Neural TTS groups participants perceived the factor as significantly more important compared with Standard TTS (U Mann-Whitney test, $U(74) = 2.397, p = .017$). However, due to unequal sample sizes in both conditions ($N=28$ in \emph{control} vs. $N=73$ in \emph{intervention}), 
we suggest interpreting this result with caution.

\begin{table*}[h]
\caption{Related-Samples Friedman's Two-Way Analysis of Variance by Ranks between users perception of the importance of parameters for user's choices in the task. $p$-values are Bonferroni-corrected, to guard against multiple comparisons.}
\begin{tabular}{llll}
\toprule
Omnibus test $\chi^2(6, N = 73) = 80.273, p < .001$ \\ \midrule
\textbf{Pairwise Comparisons}                              & \textbf{Z-statistic}    & \textbf{Adj. $\bm{p}$-value} \\ \midrule
Presentation
factors vs. Health Considerations        & -.287              & 1.000                \\
Presentation
factors vs. Novelty                      & -1.092             & 1.000                \\
Presentation
factors vs. Combination of Ingredients   & -2.241             & .525                 \\
Presentation
factors vs. Familiarity                  & -3.774             & .003                 \\
Presentation
factors vs. Previous Positive Experience & -3.908             & .002                 \\
Presentation
factors vs. Ingredients                  & -6.398             & $<$ .001      \\ \bottomrule
\end{tabular}
\label{tab:varbyrank}
\end{table*}

We also ran a regression analysis to assess the relationship between the number of choices favouring high-fidelity voice presentation (dependent variable) and two predictors: ``Presentation Factors'' and ``The Impact of Presenting on Decisions.'' The results ($R^2 = .057, F(2,73) = 2.129, p = .127$) showed that neither of these predictors had a significant impact on the dependent variable.

Additionally, a non-parametric correlation analysis (Spearman's $\rho$) also did not reveal significant correlation between ``Presentation Factor'' - ``N of choices towards Neural TTS'' ($\rho(99) = .199, p = .091$ and ``Impact of Voice on Decisions'' - ``N of choices towards Neural TTS''  ($\rho(99) = -.024, p = .843$. See \autoref{tab:rho} for details.

\begin{table*}[h!]
\caption{Sperman's $\rho$ (non-parametrical) correlation model.}
\begin{tabular}{lllll}
\toprule
\multicolumn{1}{c}{} \textbf{N of choices towards NeuralTTS voices} &
  \multicolumn{1}{c}{\textbf{Sperman's $\rho$}} &
  \multicolumn{1}{c}{\begin{tabular}[c]{@{}c@{}}\textbf{$\bm{p}$-value}\end{tabular}} &
  \multicolumn{2}{c}{\textbf{95\% CI}} \\ \midrule
\begin{tabular}[c]{@{}l@{}} Presentation Factors\end{tabular} &
  .199 &
  .091 &
  [-.039 &
  .416] \\ 
\begin{tabular}[c]{@{}l@{}} The Impact of Presenting on Decisions\end{tabular} &
  -.024 &
  .843 &
  [-.259 &
  .214] \\ \bottomrule
\end{tabular}
\label{tab:rho}
\end{table*}

\section{Discussion}\label{discussion}
We have examined the relationship between perceived characteristics of synthetic voice (i.e., ease of understanding, listening enjoyment, and domain suitability) and its impact on user behaviour. 
Our study provides an indication that, when provided with two options, presented by two different synthetic voices, users are more likely to pick one that is provided by a high-fidelity, neural synthetic voice rather than a standard concatenative voice. Interestingly, users seem unaware of the impact that voice characteristics may have on their decisions, and consider it as the least influential factor for their decisions. This result indicates existence of a potential dark pattern in the voice interfaces that can be exploited to steer users' decisions in scenarios that feature multiple voices. Below, we provide answers to our RQs and discuss some implications for use and design of interfaces that feature synthetic speech. 

With respect to our \textbf{RQ1:} \textit{\textbf{``To what extent the presentation of options provided with synthetic voices can affect users’ choices?''}}, our study suggests that when given a choice between two options presented by two different synthetic voices, users are more inclined to select the one provided by a high-fidelity, neural synthetic voice over a standard concatenative voice. This effect is deemed to have a small-to-medium effect size (cf.~\autoref{sec:behaviour}). However, the absence of significant differences between two high-fidelity voices (cf. ~\autoref{tab:Krus}) suggests that the specific parameters causing this effect remain unidentified. As was shown by the results of Stage 1, users perceived all three parameters combined; it is reasonable to assume that if users like the voice, they are not reflecting much about its specific parameters. In the same manner, previous studies showed connection between parameters of listening enjoyment and perceived trustworthiness of the voices~\cite{belin2017sound,mcaleer2014you}. 

Combining the results of Stages 1 and 2, we infer that subjective metrics of enjoyment and/or domain suitability might be decisive predictors of voice influence. Specifically, if the people like the voice, they may be more willing to follow it without reflecting on it. Yet, the exact prosodic correlates remain an open question. One possible explanation of the found effect lays in the differences in perception of our Standard TTS voice which, due to its slower pace, it could have made recognition and recall of information more challenging~\cite{rodero2016influence,rodero2022expressive}, compared to both Neural TTS voices that are faster (cf.~\autoref{fig:pros}). While the mean of understandability for this voice was above neutral score (3/5), and the extensive commercial use of this voice in the last ten years makes it unreasonable to think that participants did not understand the options provided by the voice. Moreover, it is also possible that higher pitch variance of our Neural voices which makes speech perceived as more dynamic and thus more attractive~\cite{scherer2003vocal}, led participants to select options provided by these voices more often.

On the other hand, in response to \textbf{RQ2:} \textit{\textbf{``How impactful do users consider the role of voice in affecting their choices?''}}, we received mixed results regarding users' awareness of how voice presentation influenced their choices. Although there is evidence suggesting the increased importance of presentation in the experimental group compared to the control, both groups ranked presentation factors as least important among all considered factors (see  ~\autoref{tab:varbyrank} and ~\autoref{fig:influ}). This implies that even if we cannot definitively state that users are oblivious to the influence of voice presentation, they might underestimate its potential impact on their decisions. The effect observed might be explained by the theory of ``Cognitive Dissonance''~\cite{harmon2019introduction}. This theory suggests that individuals naturally want to maintain a consistent self-perception and congruence in their decisions. Essentially, when people are questioned about their choices in a well-known domain, like food, they often lean on past experiences and the reasons that they have previously used to justify similar decisions. They subsequently use these past justifications to explain their current choices retroactively.  Notably, our discovery of identical patterns of preferences in both the experimental and control groups further supports this explanation. This consistency might indicate that individuals in both groups are resolving cognitive dissonance in similar ways, using past experiences to validate their present choices.

Our findings suggest a potential trend that could influence users' decision-making in voice-only interactions. Given that CA already showcase specific third-party features voiced by various agents~\cite{zargham2021multi}, there lies an opportunity to direct user attention towards certain services or products by making them more prominent and potentially increasing purchases. Conversely, less favourable options, such as opting out of a subscription, might be voiced in a less appealing way (e.g., by diverting user to a CA with a less attractive voice to deal with subscription management).
Furthermore, since CAs are predominantly used for low-involvement orders like takeaways that require minimal deliberation, users might not thoroughly assess their actions. Over time, this can foster detrimental habits affecting personal finances. 

In this light, the voice influence mechanism we analysed could be likened to the ``Interface Interference'' dark pattern~\cite{gray2018dark} prevalent in visual interfaces, including cookie consent banners. While in-situ studies have not yet confirmed the commercial applicability of this mechanism, stakeholders in legislation and research must be aware of such mechanisms and be proactive against their misuse.

Prior research has indicated that, although recent legislation like the Digital Service Act~\cite{eu2022/2065} aims to curb manipulative online practices, there remains a regulatory gap concerning voice-interactive virtual agents~\cite{de2023present}. The transient nature of speech adds a layer of complexity for legislators. While our findings revolve around multi-voice interaction (pairs of different voices), it is plausible that subtle modifications to a synthetic voice, such as degrading speech quality (e.g., by introducing phase inconsistencies), could influence user decisions even in single-agent interactions. This underscores the urgent need for both regulators and researchers to closely inspect this interaction mode and evaluate its potential for deception.

The recent discussions highlighted that the elements of designs that capture attention should not be immediately labelled as deceptive. Instead, their evaluation should depend on the context in which they are utilised \cite{monge2023defining}.
In this line of thought, we wish to broach the potential of leveraging voice quality effects on choices beyond the domain of dark patterns.
This can serve as a positive nudge, subtly steering user behaviour towards beneficial outcomes. For instance, the appeal of engaging voices could be harnessed in therapeutic contexts and mindful reflections, wherein expressive voices might foster positive behavioural shifts by centring users' focus on uplifting thoughts and promoting mindfulness. Nonetheless, it is crucial that any such interventions always respect user consent and agency.

\subsection{Design Implications}
Based on our findings, we have mapped the following implications for designing voice-based interactions in conversational agents:

\subsubsection{Develop Ethical Considerations and Guidelines for Voice-Interaction Design}
Our study indicates that high-fidelity neural synthetic voices can have a more pronounced influence on user decisions compared to standard voices. When designing conversational agents, caution is needed to prevent potential manipulation. Existing guidelines primarily target potential dark patterns in visual user interfaces~\cite{BEUC2022,EDPB2022}, overlooking the unique characteristics of voice interaction~\cite{de2023present}. While our research mainly examined the prosodic effects of synthetic speech on user choices, further research is needed to pinpoint voice communication aspects that may bias users. The ultimate objective is to establish guidelines for designers, ensuring that options are presented with equitable vocal attributes. We must also evaluate the accessibility of agents, especially those used commercially in multiagent contexts, to further explore their potential for intentional or unintentional influence on user choices.

\subsubsection{Encourage User Customisation and Agency}
The effects we highlighted might be more prevalent in low-stakes scenarios where users are not deeply invested in pondering their choices. A design strategy to counteract this influence is enhancing user motivation and agency when interacting with the system. One method to curtail unintended sway by synthetic voices is to permit users to adjust voice settings. By giving users the autonomy to select or modify voice traits based on their liking, designers can keep interactions user-centric. Alternatively, the system could offer more incentives for users to reassess their decisions, promoting deliberate and user-centred choices~\cite{dubiel2023you,danry2023don}.

\subsubsection{Define Domain Parameters for Tailored Interaction}
Our study suggests that domain suitability can be crucial when crafting voice interactions. Analysing voice interaction features that resonate with human-to-human communication within specific domains can enhance user experience. This can pave the way for the creation of specialised agents applicable in both multiagent and domain-specific contexts both for transactional ~\cite{zargham2021multi}  and social ~\cite{sutton2019voice,aylett2020voice} types of interactions. 

\section{Limitations and Future work}
We are mindful that our study is subject to several limitations. 
It should be noted that our experiment was limited to a one-off interaction. Therefore, different results might have been observed with repeated exposure to different voices. In the presented experiment, we modelled a ``low-involvement interaction'' scenario, where the user's motivation to choose a specific option is relatively low. We did not expect the effect to appear in scenarios where users clearly prefer one option over others beforehand.

We would like to note that our findings should be considered in the context of a `food inspirations' scenario and may not generalise beyond this domain.
We also acknowledge that the results regarding the user's awareness of voice manipulation may be influenced by the manner in which we posed questions about the importance of this factor. We summarised the voice effects as ``the way the assistant presented the menu options,'' which could potentially affect the users' broader interpretation of the question. Although our pretests indicated that users typically interpreted the question in terms of voice parameters (such as intonation, pronunciation quality, pleasantness of voice, absence of artifacts, etc.), further studies are required to precisely determine the extent of users' awareness of voice quality/prosodics manipulation.

In the future we to plan experiment with other domains and to expand our experiments to female voices to see if the behavioural effect persists,
since the literature highlights the fact that the gender dimension is one of the important in the voice perception domain~\cite{mullennix2003social}.

\section{Conclusion}\label{conclusion}
While many studies have explored potential scenarios in which voice interfaces could be manipulated to distort user choices, to the best of our knowledge, this work is the first to highlight the possibility of ``interface interference'' dark patterns in voice interfaces. Such patterns, which are evident in visual interfaces, can guide users toward choices favoured by company or a service provider. Just as the different colouring of ``Accept'' and ``Reject'' buttons on cookie consent forms on websites can influence users' decisions, our research suggests that voice assistants might use deceptive strategies by audibly differentiating choice options. We found that users often underestimate the influence of voice on their choices. As multi-agent communication evolves, there is an opportunity to exploit this, directing users toward choices based on their subconscious attraction to a specific agent’s voice. Moreover, with the growing development of voice-based technologies in everyday life, we should become more vigilant about their manipulative potential.
\begin{acks}
This work is supported by the Horizon 2020 FET program of the European Union 
through the ERA-NET Cofund funding (BANANA, grant CHIST-ERA-20-BCI-001)
and Horizon Europe's European Innovation Council 
through the Pathfinder program (SYMBIOTIK, grant 101071147). This work is also supported by the Luxembourg National Research Fund
(FNR) Decepticon (grant no.IS/14717072).
\end{acks}

\bibliographystyle{ACM-Reference-Format}


\onecolumn
\appendix

\section{Appendix}\label{appendix}

\subsection{Prompts used in our study}

Pair 1:
\begin{itemize}
    \item 1a. Grilled chicken with roasted vegetables and quinoa pilaf.
    \item 1b. Grilled cod with steamed asparagus and wild rice medley.
\end{itemize}

Pair 2:
\begin{itemize}
    \item 2a. Creamy pasta with sautéed garlic, mushrooms, spinach, and Parmesan cheese.
    \item 2b. Flavorful pasta with rich tomato sauce, Italian herbs, and grated Parmesan.
\end{itemize}

Pair 3:
\begin{itemize}
    \item 3a. Nourishing snack with sliced apples, almond butter, and a sprinkle of cinnamon.
    \item 3b. Wholesome snack with carrot sticks, hummus, and crunchy whole-grain crackers.
\end{itemize}

Pair 4:
\begin{itemize}
    \item 4a: Oat porridge with honey and figs
    \item 4b: Croissant with almond butter and berries.
\end{itemize}

\subsection{Voice impact study, Prolific protocol:}

\textbf{Page 1}: 
Thank you for joining our study on factors that inspire users' choices of new recipes provided by voice assistants. 
Please follow these steps:
\begin{itemize}
    \item Please read the consent form and enter your Prolific ID in the consent form on the next page if you are agree to participate.
    \item In each of the following questions, you will be presented with four pairs of meal options. Please choose the one from each pair that seems more appealing to you.
    \item Ensure your headphones are on, as options will be presented via voice assistants like Alexa or Siri.
    \item After making selections for four pairs of meals, you will be given questionnaires about 
    the reasons for your choices.
\end{itemize}

\textbf{Page 2}:
Consent form for the study provided by the University of Luxembourg.

\textbf{Page 3}:
This page displays four pairs of options, each spoken by different voices based on the participant's assigned condition. Participants can replay the options if needed. For each pair, participants must select one option in response to the question: ``Which option do you find more appealing?''

\textbf{Page 4}: 

(I) Please rate impact of the following factors on meal options that you have chosen in this study for 5-point scale (from ``Not Influential at all'' to ``Really Influential'')

\begin{itemize}
    \item \textbf{Ingredients}: specific ingredients that caught your attention or appealed to your taste.
    \item \textbf{A novel combination of the ingredients}: unique pairings or creative blending of ingredients that intrigued you.
    \item \textbf{Health and nutritional considerations}: the perceived healthiness or nutritional value of the meal.
    \item \textbf{Familiarity with the ingredients or cuisine}: preference for ingredients or dishes that you are familiar with.
    \item \textbf{Previous positive experiences with similar dishes}: positive memories or past enjoyment of similar meals.
    \item \textbf{Variety or novelty}: desire to try something new or different from your usual choices.
    \item \textbf{ Presentation factors}: the way in which the  assistant presented the options.
\end{itemize}

(II) Please assess the following statements, considering the Voice Assistant presentation of the meal options: 

\begin{itemize}
    \item I think that Voice Assistant was useful in presenting meal options.
    \item I think that the style of presenting choices by the Voice Assistant is important for making decisions.
\end{itemize}

\subsection{Results of analysis of gender effects on the voice perception characteristics}
The summary of results is presented in \autoref{tab:gender}.

\subsection{Means and Confidence interval of each of the parameters for each voice and factors from Stage 1}
The summary of results is presented in \autoref{tab:mean}

\subsection{Related-Samples Friedman's Two-Way Analysis of Variance by Ranks between users perception of the importance of parameters for user's choices in the task. $p$-values are Bonferroni-corrected, to guard against multiple comparisons for control condition}

The results of Related-Samples Friedman's Two-Way Analysis of Variance by Ranks between users perception of the importance of parameters for user's choices in the task are presented in \autoref{tab:varbyrankcontrol}.
\begin{table}
\caption{Summary of Independent-Samples Kruskal-Wallis Tests between parameters of the perception of each voice}
\centering
\begin{tabular}{p{9cm}ll}
  \hline
Null hypothesis (N = 50) & U Mann-Whitney & $p$-value \\\hline
{The distribution of ``Ease
  of understanding'' for Standard TTS voice is the same across categories
  of gender.} & 341.5          & .572 \\
{The distribution of ``Ease
  of understanding for Neural TTS2 voice is same across categories of gender.}         & 309.5          & .953 \\
{The distribution of ``Ease
  of understanding'' for Neural TTS1 voice is the same across categories of
  gender.}  & 316.0          & .946 \\
{The distribution of
  ``Listening Enjoyment'' for Standard TTS voice is the same across
  categories of gender.}   & 319.5          & .892 \\
{The distribution of
  ``Listening Enjoyment'' for Neural TTS2  voice is same across categories of
  gender.}       & 252.5          & .243 \\
{The distribution of
  ``Listening Enjoyment'' for Neural TTS1~voice is the same across
  categories of gender.}    & 289.5          & .653 \\
{The distribution of
  ``Domain Suitability'' for Standard TTS voice is the same across
  categories of gender.}    & 314.5          & .969 \\
{The distribution of
  ``Domain Suitability'' for Neural TTS2~~voice is same across categories of
  gender.}        & 286            & .605 \\
{The distribution of
  ``Domain Suitability'' for Neural TTS1 voice is the same across
  categories of gender.}     & 313            & .992\\ \hline
\end{tabular}
\label{tab:gender}
\end{table}

\begin{table}
\caption{Means and 95\% Confidence Intervals for Stage 1 factor evaluations for each voice (the scores are based on the total from 5 questions, ranging from 5 to 25)}
\centering
\begin{tabular}{llll}
  \hline
                      & Standard TTS               & Neural TTS1                & Neural TTS2                 \\\hline
Ease of understanding & 15.68 (95\%CI 14.42;16.94) & 19.24 (95\%CI 18.15;20.33) & 20.24 (95\%CI 19.20;21.28)  \\
Listening enjoyment   & 11.38 (95\%CI 10.12;12.64) & 16.56 (95\%CI 15.53;17.59) & 17.04 (95\%CI 15.84;18.24)  \\
Domain suitability    & 12.50 (95\%CI 11.24;13.76) & 17.42 (95\%CI 16.28;18.56) & 18.00~ (95\%CI 16.86;19.14)\\ \hline
\end{tabular}
\label{tab:mean}
\end{table}
\begin{table}[]
\caption{Related-Samples Friedman's Two-Way Analysis of Variance by Ranks between users perception of the importance of parameters for user's choices in the task. $p$-values are Bonferroni-corrected, to guard against multiple comparisons (CONTROL CONDITION)}
\begin{tabular}{llll}
\toprule
Omnibus test $\chi^2(6, N = 73) = 68.343, p < .001$ \\ \midrule
\multicolumn{4}{c}{\textit{Pairwise Comparisons (only related to “Presentation factors” question in control condition)}}                             \\
\textbf{Pairwise Comparisons}                              & \textbf{Z-statistic}    & \textbf{Adj. $\bm{p}$-value} \\ \midrule
Presentation
factors vs. Health Considerations        & -1.825              & 1.000                \\
Presentation
factors vs. Novelty                      & -.773             & 1.000                \\
Presentation
factors vs. Combination of Ingredients   & -1.825             & .525                 \\
Presentation
factors vs. Familiarity                  & -4.145             & .001                 \\
Presentation
factors vs. Previous Positive Experience & -4.237             & .001                 \\
Presentation
factors vs. Ingredients                  & -5.815             & $<$ .001      \\ \bottomrule
\end{tabular}
\label{tab:varbyrankcontrol}
\end{table}
\end{document}